\documentclass{article}
\usepackage{amsmath}
\usepackage{graphicx}
\def\p{\partial}
\def\s{\sigma}

\def\G{\Gamma}
\def\g{\gamma}

\def\d{\delta}
\def\de{\delta}
\def\D{\Delta}
\def\De{\Delta}
\def\ov{\overline}

\def\ld{\lambda}
\def\Ld{\Lambda}

\def\e{\eta}

\def\rh{\rho}

\def\b{\beta}

\def\a{\alpha}

\def\pdellx'{\frac{\partial}{\partial x'}}
\def\pdellw'{\frac{\partial}{\partial w'}}

\newcommand{\be}{\begin{equation}}
\newcommand{\ee}{\end{equation}}
\def\bed{\begin{displaymath}}
\def\eed{\end{displaymath}}
\def\bea{\begin{eqnarray}}
\def\eea{\end{eqncrray}}
\def\[{$$}
\def\]{$$}

\begin{document}

 \title{Dark Matter, Cosmic Positrons in Alpha Magnetic Spectrometer Experiment and Particle-Cosmology with Yang-Mills Gravity }
  \author{
Jong-Ping Hsu\footnote{Electronic address: jhsu@umassd.edu} and Leonardo Hsu\\
 Department of Physics, 
 University of Massachusetts Dartmouth,\\ 
North Dartmouth, MA 02747, USA\\}

\maketitle
\noindent
\ \ \ \ \ \ \ \ \ \ \ \ \ \ \ \ \ \ \ \ \ \ \ \ \ \ \ \ \ \ \ \ \ \ \     To the memory of Pauchy Hwang and T. S. Cheng   

\bigskip

\bigskip

{\small   We discuss a model of  dark matter consisting of high energy anti-electron-neutrinos with leptonic force, which is produced by the conserved leptonic charge $g_\ell$ associated with Lee-Yang's $U_1$ gauge symmetry.  Based on particle-cosmology for early universe, the high energy neutrino (HEN) model of dark matter assumes that the neutron decay processes, $n\to p^+ + e^- +\ov{\nu}_e$, dominate the epoch after the creation, collision and confinement processes of quarks and antiquarks in the beginning.  The HEN model implies the following results:  There are almost equal numbers of electrons, protons and anti-electron-neutrinos dominated the matter cosmos.  There are unobservable and ubiquitous anti-electron-neutrinos $\ov{\nu}_e$ with  leptonic charge $g_\ell$ in the universe.  Although the total mass of anti-electron-neutrino dark matter is negligible in the universe, its enhanced gravitational and leptonic forces could lead to the observed flat rotation curves due to relativistic $\ov{\nu}_e$, whose static force involves a factor $E_\nu/m_\nu\approx 10^6$.  We estimate the  leptonic charge to be $g_\ell \approx 7 \times 10^{-21}$. The model predicts that the anti-electron-neutrino dark matter  can interact with cosmic-ray protons to produce positrons, i.e. $\ov{\nu}_e + p^+ \to e^+ +n$, through weak interaction of the unified electroweak theory.  The anti-electron-neutrino dark matter sheds light on the Alpha Magnetic Spectrometer (AMS) experiment, which has detected the intriguing excess of cosmic-ray positrons over what is expected.  The HEN model of dark matter suggests an experimental test of the new Lee-Yang force between electrons by using modern precision Cavendish experiment. }

\bigskip 

\bigskip

\section{Introduction}
Based on particle-cosmology in flat space-time, we discuss a model of anti-electron-neutrino  (anti-e-neutrino) dark matter.  Cosmic-scale phenomena, such as the anomalous galactic rotation curves, that are currently attributed to a gravitational interaction with an as-yet-unobserved ``dark matter" are actually the results of an enhanced Yang-Mills gravity for high energy neutrinos and an interaction between anti-e-neutrinos $\ov{\nu}_e$ and ordinary electrons in stars.  The interaction between these leptons is a hitherto unobserved force associated with the established conservation of leptonic charge.

 In 1955, Lee and Yang discussed a  local $U_1$ gauge symmetry associated with the conservation of baryon charge (or number)\cite{1,2}.  They showed that the dynamic equations of the baryonic $U_1$ gauge field are formally the same as Maxwell's equations.  Based on $E\ddot{o}tv\ddot{o}s$'s experiment\cite{1}, Lee and Yang estimated the new baryonic force to be weaker than the gravitational force by a factor of $10^{-5}$ or smaller.

However, Lee and Yang's idea and discussions can also be applied to the conserved leptonic charge, so that there should be an inverse-square leptonic force between leptons, in addition to the long range Coulomb force and the short range weak force.  It appears that the baryonic force of protons and neutrons associated with baryon charge is  too weak to affect the rotation curves of spiral galaxies.  Moreover, it is unlikely that baryons can play the role of invisible dark matter because they have strong and electromagnetic interactions.  In contrast, neutrinos do not have these problems.
Because such a leptonic Lee-Yang force and its $U_1$ gauge field have not been detected in high energy experiments, some physicists consider the $U_1$ symmetry corresponding to the leptonic charge not to be a local gauge symmetry.  However, particle physics strongly suggests the primacy of the gauge symmetry principle in the physical world\cite{2,3,4}, which is supported by the success of QED, QCD and the unified electroweak theory.  Although this as-yet-undetected leptonic force is very weak, it could  play a role in cosmic phenomena involving an extremely large number of leptons.  Thus, it is natural to postulate the principle of universal gauge symmetry:

 {\em  All conserved charges are associated with local internal gauge symmetries.\footnote{To include gravity, the local external space-time translation gauge symmetry for quantum Yang-Mills gravity\cite{4} should be incorporated.  These gauge symmetries are dynamic symmetries.\cite{4} }}
 
In this paper, we will discuss how this principle of universal gauge symmetry and the HEN model of dark matter can be tested by experiments involving cosmic-scale phenomena, e.g., the flat rotation curve of galaxies and Alpha Magnetic Spectrometer (AMS) experiments.

Let us first examine briefly how this model of anti-e-neutrino dark matter taking the Lee-Yang force between leptonic charges can explain the observed rotation curve of galaxies. A rotation curve is obtained by measuring the speed of stars in a galaxy as a function of their distances from the center of the galaxy. The measured rotation curve can be compared with an expected rotation curve, which is obtained using visible stars to estimate the mass distribution of the galaxy and Newton's law of universal gravitation.  As is well known, the motion of stars in galaxies do not correspond to the expected rotation curve based on the visible stars and gas and this discrepancy is usually attributed to the presence of dark matter with gravitational interaction.  However, as we shall see the $U_1$ gauge symmetry for the leptonic charge $g_{\ell}$ produces a leptonic Lee-Yang force  $F^{LY}=-g_{\ell1} g_{\ell2}/(4\pi r^2)$ that is also an inverse-square law force, in the usual non-relativistic static limit.  We demonstrate that the combination of the enhanced Yang-Mills gravity (section 4) and the Lee-Yang force (section 5) results in the observed flat rotation curve.  

  Although it is often stated that the dark matter problem is based on a well-confirmed theoretical construction, our present understanding of the universe is, in general, extremely limited.  Thus, it is reasonable to investigate cosmology on the basis of particle-cosmology, especially for phenomena of the early universe.   Our discussions in this paper are based on the laws and properties established by theoretical particle physics and experiments which, in itself, comprise a well-confirmed theoretical construction.  If a model has little to do with particle physics and/or violates established symmetry principles and conservation laws, it is probably too speculative. 

\section{Particle-cosmology and the creation of anti-e-neutrino dark matter}

This model of anti-e-neutrino dark matter is supported by established conservation laws such as lepton and baryon charges, etc. and the observed processes in particle physics. Crucial to this model is an assumption that in the epoch immediately following the creation of quarks and antiquarks at the beginning of the universe,\footnote{Since neutrinos and leptons do not interact via the strong force, they would have been directly produced in negligible numbers compared to quarks and hadrons.} the neutron decay process, $n\to p^+ +e^- +\ov{\nu}_e$, was the dominant process. As a result, these three stable particles should be roughly equal in their numbers and dominate the observable `matter cosmos.'  This result is consistent with experiments that the observable cosmos is electrically neutral and dominated by the hydrogen atom.

The typical picture of the beginning of the universe is similar to that inside a gigantic super-collider. The typical strong interaction time $\tau_{int}$ between two quarks is about the same or shorter than that of two hadrons in a high energy laboratory, i.e., roughly $(1 fermi)/(speed \ of \ light) \approx 10^{-14} s\approx \tau_{int}$, so the time interval between the creation of quarks and antiquarks and the appearance of hadrons\footnote{Baryons and mesons that participate in strong interactions.} is presumably extremely short. The subsequent collision and confinement processes of quarks to produce 3-quark bound states (baryons) and quark-antiquark bound states (mesons) would then take on the order of $\approx 10^{-10}s$, presumably. 

The assumption that the neutron decay process was dominant during this epoch is equivalent to assuming the dominance of the formation of neutrons and N resonances, etc.\cite{5} during this confinement processes of quarks and antiquarks.  The reason is that these N resonances, etc. will participate in strong decay processes (e.g., a neutron resonance $\to$ neutron and meson or mesons) within, say, the interaction time interval $\tau_{int}\approx 10^{-14} s$.  The physical reason for this estimation is that the mean lifetimes of N resonances are about $10^{-24} s$, which corresponds to the Brett-Wigner full width $\approx 300 MeV$\cite{5}. Thus, N resonances would immediately become neutrons, which would then decay via the weak interaction with a `very long' mean lifetime, roughly $\tau_n= 885.7 s >> 10^{-24} s$\cite{5}. According to the exponential decay law, about $t_s\approx 2\times \tau_n$ after the creation of quarks and antiquarks, most of the neutrons and N resonances (i.e., about $ 87\%$) would have decayed to form equal numbers of protons, electrons and anti-e-neutrinos. Roughly speaking, this time $t_s$ sets the time scale for other process such as nucleosynthesis to take place.  Thus, the ubiquitous and unobservable anti-e-neutrinos dark matter are created near the beginning of the universe and affect the ensuing evolution, according to this model based on particle-cosmology.

Of course, there are many other non-dominant decay processes that lead to comparatively tiny non-equal numbers of protons and electrons.  Because of the relatively long lifetime of the neutron, the time of nucleosynthesis in this model of dark matter is estimated to be much later than that in the standard big bang model.  However, this model is consistent with the fact that there are almost equal numbers of electrons and protons in the observable cosmos.

To understand the anti-e-neutrino dark matter, let us briefly discuss the framework of particle-cosmology\cite{7}.  Particle physics and quantum field theories are closely related to the Poincar\'e transformations (or the Poincar\'e group) in flat space-time.  But general relativity is incompatible with the Poincar\'e group.  Thus, there is a conceptual problem for particle-cosmology if general relativity is used for cosmology.  Dyson\cite{4} stressed that `The most glaring incompatibility of concepts in contemporary physics is that between Einstein's principle of general coordinate invariance and all the modern schemes for a quantum-mechanical description of nature.'  However,  quantum Yang-Mills gravity has been formulated on the basis of  translation gauge symmetry in flat space-time and is consistent with experiments\cite{4,8,9}. 

In the geometric-optics limit, the quantum wave equation of the fermion field in Yang-Mills gravity reduces to a Hamilton-Jacobi type equation $G^{\mu\nu}(\p_\mu S)(\p_\nu S) - m^2=0$, where $G^{\mu\nu}=\e_{\a\b}J^{\a\mu} J^{\b\nu}$\cite{4}. This equation of motion for macroscopic objects in flat space-time involves a new effective Riemannian metric tensor $G^{\mu\nu}$.  Classical objects move in Yang-Mills gravitational field as if they were in a curved space-time.  Thus, the apparent curvature of space-time appears to be simply a manifestation of the flat space-time translational gauge symmetry for the motion of quantum particle in the classical limit.  It turns out to be formally the same  as the corresponding equation for macroscopic objects in general relativity and is called `Einstein-Grossmann' equation of motion\cite{4}.  This equation is crucial for Yang-Mills gravity to be consistent with the perihelion shift of the Mercury, the deflection of light by the sun and the equivalence principle\cite{4}.   Yang-Mills gravity gives an elegant explanation as to why gravitational force is always attractive between electron-electron and between electron-positron, in contrast to leptonic Lee-Yang force and electromagnetic force, as shown in equations (9)-(11) below.  The combination of gauge symmetry and an effective metric tensor provides a framework for and leads to alternative dynamics of cosmic expansion based on Yang-Mills gravity at the super-macroscopic limit.  In this super-macroscopic limit  with the cosmological principle, one has cosmic Okubo equation for the motion of distant galaxies and for cosmic red shifts, which are consistent with experiments\cite{7}.

There is no conceptual problems for particle-cosmology based on Yang-Mills gravity.  The reason is that Yang-Mills gravity is based on translation gauge symmetry in a flat space-time with inertial frames and can be quantized\cite{4}.  Thus, one now has a comprehensive particle-cosmology that includes quantum Yang-Mills gravity, particle physics and quantum field theories. 

 The high energy neutrino (HEN) model demonstrates that anti-e-neutrinos are a viable dark matter candidate and deserve further investigation.  This model represents an alternative approach to understanding dark matter phenomena that does not depend on the existence of as-yet undiscovered particles and other speculations. The key features of the HEN model are as follows:  (i) It provides an understanding of the flat galactic rotation curves without drastically increasing the total observed mass of the universe by a factor of 5 to 60, in contrast to all other interpretations of dark matter.  (ii) It is based on the well-established principle of gauge symmetry for quantum Yang-Mills gravity.  (In particular, at the super-macroscopic limit, Yang-Mills gravity and the cosmological principle lead to a Hamilton-Jacobi type equation for the motion of galaxies.  Its solutions are consistent with Hubble's Law and cosmic redshifts.)   (iii) It is based on particle-cosmology in a flat space-time with CPT invariance, implying the conservation of leptonic and baryonic charges, as well as an electrically neutral universe.   (iv) The HEN model predicts that the interaction between these ubiquitous anti-e-neutrinos and cosmic-ray protons leads to an excess of cosmic-ray positrons, which have been detected by the PAMELA experiment and the Alpha Magnetic Spectrometer experiment.
 
\section{Lagrangian for leptonic $U_1$ gauge field and the attractive force of Yang-Mills gravity }
Let us now make the discussion quantitative and include both leptonic and (Yang-Mills) gravitational interactions in the Lagrangian $L_{\ell\phi}$\cite{7,8} within the framework of particle-cosmology.  In an inertial frame, the Lagrangian $L_{\ell\phi}$ involving a set of leptons $\ell=(e, \nu_e)$, the leptonic $U_1$ gauge field $L_\mu(x)$, and gravitational spin-2 fields\cite{4} $\phi^{\mu\nu}(x)$ takes the form,
\be
S_{\ell\phi}= \int L_{\ell\phi} d^4x,  \ \ \ \ \ \ \ \ c=\hbar=1,
\ee
$$
L_{\ell\phi}= i[\overline{\ell} \g_\mu \Delta^\mu \ell ] -
m_{\ell}\overline{\ell} \ell -\frac{1}{4}L_{\mu\nu} L^{\mu\nu},   
 $$ 
 \be
+ \frac{1}{2g^2}\left (C_{\mu\a\b}C^{\mu\b\a}-
 C_{\mu\a}^{ \ \ \  \a}C^{\mu\b}_{ \ \ \  \b} \right), \ \ \ \ \ \   g^2=8\pi G,
\ee
$$
C^{\mu\nu\ld}=J^{\mu\s}\p_\s J^{\nu\ld}- J^{\nu\s} \p_\s J^{\mu\ld}, \ \ \ \  J^{\mu\nu}=\e^{\mu\nu}+
g \phi^{\mu\nu} = J^{\nu\mu},
$$
\be
\Delta^\mu \ell = (\p^{\mu}+
g \phi^{\mu\nu} \p_\nu -ig_{\ell} L^\mu) \ell,  \ \ \ \  L_{\mu\nu}=\p_\mu L_\nu-\p_\nu L_\mu,
\ee
where $C_{\mu\a\b}C^{\mu\b\a}=(1/2)C_{\mu\a\b}C^{\mu\a\b}$, $G$ is the Newtonian constant and $\e^{\mu\nu}=(1,-1,-1,-1)$.  The gauge-fixing terms\cite{4} for $\phi^{\mu\nu}$ and $L_\mu$ are not included because we do not consider quantization of fields here.

From the gauge invariant action (1), one can derive equations for the leptonic gauge fields $L_\mu(x)$ and tensor field $\phi^{\mu\nu}(x)$, 
\be
 \p^\mu L_{\mu\nu} + {g_\ell} \ov{\ell} \g_\nu  \ell = 0,
\ee
\be
H^{\mu\nu} =  g^2 S^{\mu\nu},
\ee
$$
H^{\mu\nu} = \p_\ld (J^{\ld}_\rho C^{\rho\mu\nu} - J^\ld_\a 
C^{\a\b}_{ \ \ \ \b}\e^{\mu\nu} + C^{\mu\b}_{ \ \ \ \b} J^{\nu\ld})  
$$
\be
- C^{\mu\a\b}\p^\nu J_{\a\b} + C^{\mu\b}_{ \ \ \ \b} \p^\nu J^\a_\a -
 C^{\ld \b}_{ \ \ \ \b}\p^\nu J^\mu _\ld,
\ee
 where the `source tensor' $S^{\mu\nu}$ of fermion matter is given by
\be
S^{\mu\nu} =  \overline{\ell} i\gamma^\mu \p^\nu \ell . 
\ee
The indices $\mu$ and $\nu$ in (5) should be made symmetric\cite{4}.  The gauge fields $L_\mu$  are generated by leptonic charges.  The equation of lepton is given by
\be
\left[i\g_\mu (\p^{\mu}+g\phi^{\mu\nu} \p_\nu  - i g_{\ell}{L^{\mu}}) - m_{\ell}\right] \ell = 0. 
\ee

Gravity has been discussed from the viewpoint of Yang-Mills theory and its generalization\cite{8,9}.
Suppose one assumes the external space-time translation gauge group.  Its gauge covariant derivative will be very similar to that of electrodynamics with $U_1$ gauge symmetry, and enables us 
to see the attractive nature of gravity explicitly.   Suppose we include the electromagnetic  interaction in the wave equation of lepton (8).  Then, the total gauge covariant derivative associated with the electron (i.e., charged lepton) and  its complex conjugate associated with the positron are respectively given by,
\be
\p^{\mu}+g\phi^{\mu\nu} \p_\nu  - i g_{\ell}{L^{\mu}} -ie A^\mu,
\ee
and
$$
\p^{\mu}+g\phi^{\mu\nu} \p_\nu  + i g_{\ell}{L^{\mu}} +ie A^\mu.
$$
The basic properties of the electric and leptonic (Lee-Yang) forces, say, between electron-electron, $[e^-, e^-]$, and between electron-positron $[e^-, e^+]$ are
\be
[e^-, e^-]:  \    (-ie)(-ie)=-e^2, \ \ \ \ \   (-ig_{\ell})(-ig_{\ell})=-g^2_{\ell},  \ \ \ \   repulsion,
\ee
$$
[e^-, e^+]:  \    (-ie)(+ie)=+e^2, \ \ \ \ \   (-ig_{\ell})(+ig_{\ell})=+g^2_{\ell},  \ \ \   attraction,
$$
Thus, we have repulsive (and attractive) electric and leptonic  forces, which are due to the presence of $i$ in their couplings with $U_1$ gauge symmetry.  In contrast, the Yang-Mills gravitational forces $F_{YMg}$ between electron-electron and between electron-positron are always attractive,
\be
F_{YMg}[e^-, e^-]:  \ \ \    (g)(g)=+g^2,  \ \ \   attraction,
\ee
$$
F_{YMg}[e^-, e^+]:  \ \ \    (g)(g)=+g^2, \ \ \    attraction.
$$
The reason is that the gravitational couplings in the gauge covariant derivatives in (3) do not involve $i$ because the space-time translation ($T_4$) group dictates the coupling term in the gauge covariant derivative is $\p^{\mu}-ig\phi^{\mu\nu} p_\nu,$  where a representation of the generator of the $T_4$ group is $p_\mu=i\p_\mu$.

  \section{Relativistic and non-relativistic approximations of static solution of gravitational potentials}
  
  As usual, in the non-relativistic static limit we make the static approximation by the replacement,
$ source \to  \de^3({\bf r})$, for the point source particle at rest at the origin.  However, this is not be a suitable approximation for particles with very small masses such as neutrinos with energies much than their masses.  For example, suppose one considers the `gravitational force' between, say, a relativistic particle and a macroscopic body is determined by their energy-momentum tensor in general relativity.  For the force between a relativistic electron (or a photon) with energy $E$ and velocity ${\mbox{\boldmath $ \b$}}={\bf v}/c$ and a macroscopic body with a big mass M is 
\be
{\bf F}_g = \frac{-GM E[{\bf r}(1+\b^2)-{\mbox{\boldmath $ \b$}}({\mbox{\boldmath{$ \b$}}} \cdot {\bf r})]}{r^3},\ \ \ \ \ \       c=\hbar = 1,
\ee
which  was obtained by Okun based on general relativity\cite{10}.   The result (12) reduces to the usual non-relativistic static limit when $E\to m$ or $\b\to 0$. 

In this model of dark matter, let us consider general static approximations for gravitational potential in flat space-time based on Yang-Mills gravity.  In inertial frames with the metric $ \eta_{\mu\nu}=(1,-1,-1,-1)$, it suffices to consider the gauge-field equation (5) with linearized wave equation,
$$
\p_\ld \p^\ld \phi^{\mu\nu} -  \p^\mu \p_\ld \phi^{\ld\nu} -
\eta^{\mu\nu} \p_\ld \p^\ld \phi  + \eta^{\mu\nu} \p_\a \p_\b \phi^{\a\b}
$$
\be
+  \p^\mu \p^\nu \phi - \p^\nu \p_\ld \phi^{\ld\mu} - g S^{\mu\nu} = 0.
\ee
The source $g S^{\mu\nu}$ by itself cannot generate the (00) component of field $\phi^{00}$.  Nevertheless, this equation can be written in the following form\cite{4}:
\be
\p_\ld \p^\ld \phi^{\mu\nu} - \p^\mu \p_\ld \phi^{\ld\nu} +
 \p^\mu \p^\nu \phi^\ld_\ld - \p^\nu \p_\ld \phi^{\ld\mu} = g (S^{\mu\nu}
- \frac{1}{2}\eta^{\mu\nu}
 S^\ld_\ld).
\ee
  It is interesting that the linearized gauge-field equation (14) is mathematically the same as the
corresponding linearized equation in general relativity.  The $\phi^{00}$ component in (14) lead to the gravitational potential in Yang-Mills gravity\cite{4}.

For the relativistic static approximation, we note that the source term $S^{\mu\nu}$ in (7) transforms the same as the energy-momentum tensor $T^{\mu\nu}$ of a classical particle.  Thus, we express the source as follows:
\be
g (S^{\mu\nu}
- \frac{1}{2}\eta^{\mu\nu} S^\ld_\ld) \approx  g (T^{\mu\nu}
- \frac{1}{2}\eta^{\mu\nu} T^\ld_\ld).
\ee
In analogy to charge density, one usually writes\cite{11}
\be
(A): \ \ \ \ T^{\mu\nu}=\rho \frac{dx^\mu}{ds}\frac{dx^\nu}{dt},   \ \ \ \  ds=\sqrt{1-\b^2} \ dt,  
\ee
$$
(B): \ \ \ \ T^{\mu\nu}=\rho_o \frac{dx^\mu}{ds}\frac{dx^\nu}{ds}, \ \ \ \   \rho_o=\rho \sqrt{1-\b^2}.
$$
  We also have  
\be
T^{\ld}_\ld= \frac{\rh E}{m} (1- \b^2), \ \ \ \ \    E/m=1/\sqrt{1-\b^2}.
\ee

From (15)-(17), we could make two different high energy approximations for the source tensor $T^{\mu\nu}$ to the non-Lorentz invariant static limits:
\be
(A): \    \rho \to  m\de^3({\bf r}), \ \ \    T^{00}=\rho \frac{dx^0}{ds}\frac{dx^0}{dt} \to m(E/m)\de^3({\bf r}), \ \ \ \ 
\ee
\be
(B):  \  \rho_{o} \to  m\de^3({\bf r}), \ \ \    T^{00}=\rho_o \frac{dx^0}{ds}\frac{dx^0}{ds} \to m(E/m)^2\de^3({\bf r}),   
\ee
where $t=x^0$.  Note that both the approximations $T^{00}$ in (18) and (19) reduces to the usual non-relativistic static limit $T^{00} \to m\de({\bf r}) $ when $E \to m$.  So far, there is no experiment to test these two different approximations.  Thus, we shall consider both of them to explore their comic implications.
Based on (14)-(19), we have
\be
 \nabla^2 \phi^{00}=-(1/2) g E (1+\b^2) \de^3({\bf r}),  \ \ for \ \  \rho \to  m\de^3({\bf r});
\ee
$$
\nabla^2 \phi^{00}=-(1/2) g E(E/m) (1+\b^2) \de^3({\bf r}),  \ \ for \ \  \rho_o \to  m\de^3({\bf r}).
$$

In Yang-Mills gravity based on translational gauge symmetry flat space-time, the motion of a classical particle is described by the Hamilton-Jacobi type equation\cite{4} 
\be
G^{\mu\nu}(\p_\mu S)(\p_\nu S) - m^2=0, \ \ \       G^{\mu\nu}=\e_{\a\b}(\e^{\a\mu} +g\phi^{\a\mu}) (\e^{\b\nu}+g\phi^{\b\nu}). 
\ee
 This equation is derived from the electron wave equations in the geometric-optics (or classical) limit. The classical equation (21) involves an effective metric tensor $G^{\mu\nu}(x)$, which depends on the tensor fields $\phi^{\mu\nu}$ and is  formally the same as the corresponding equation in general relativity.\cite{4,12}   In the classical limits, $\phi^{00}$ plays the role of the static Newtonian potential energy, equations in (20) lead to
 \be
 \phi^{00} = (1+\b^2)\frac{gE}{ 8\pi r}, \ \ \ \    \ \ for \ \  \rho \to  m\de^3({\bf r}),
 \ee
 $$
  \phi^{00} = (1+\b^2)\frac{gE(E/m)}{ 8\pi  r}, \ \ \ \    \ \ for \ \  \rho_o \to  m\de^3({\bf r}).
 $$
  Thus, the enhanced Yang-Mills gravitational forces $(-d/dr)[-M'g\phi^{00}]$  between a relativistic neutrino (or anti-neutrino) with energy $E=E_\nu$ (mass $m=m_\nu$) and a macroscopic body with mass $M'$  are given by
  \be
{ F}_{YM} = \frac{-GM' E_\nu(1+\b^2)}{r^2},     \ \ \ \   \ \ for \ \  \rho \to  m_\nu \de^3({\bf r});
\ee
$$
{ F}_{YM} = \frac{-GM' E_\nu(E_\nu/m_\nu)(1+\b^2)}{r^2},     \ \ \ \   \ \ for \ \  \rho_o \to  m_\nu \de^3({\bf r});
$$
where $m_\nu$ is the neutrino mass, and G is Newtonian constant, $G=g^2/8\pi $.  The relativistic static forces can be applied to the present model of anti-e-neutrino dark matter to understand the flat rotation curve.   The usual gravitational force between a neutrino at rest and a star is negligible because the neutrino mass is extremely small, i.e., $m_\nu < 1.1 \ eV$ (tritium decay).  However, the static gravitational force for a relativistic neutrino with $E_\nu > 1 MeV$ and a star with mass $M'$, the gravitational force (23) of neutrinos is no longer negligible because one could have, e.g., $E_\nu/m_\nu > 10^6$, for cosmic anti-e-neutrinos in the model.
  
 \section{Relativistic approximations of static solution of leptonic force}

Let us consider the relativistic static potential for the leptonic Lee-Yang force.
 In the static limit, where $L_\mu = L_\mu({ r})$, equation (4) with $\mu=0$ can be written as 
\be
[ \p^2 L_\mu - \p_\mu \p^\nu L_\nu]_{\mu=o} = -\nabla^2 L_0({ r}) = - {g_\ell} \ov{\ell}\g_0 \ell, 
\ee
where the relativistic neutrino source term in (24), i.e., $\ell=\nu$, transforms like the zeroth component of a 4-vector, so that we have the static approximation, $ \ov{\ell}\g_0 \ell\to  (E_\nu/m_\nu)\de^3(\bf r)$\cite{11,12}.

  The zeroth component static gauge potential $L_\mu({ r}),  \mu=0$, then satisfies the equation in such a `relativistic static limit',
\be
 \nabla^2 L_0 ({ r}) \approx  {g_\ell}(E_\nu/m_\nu) \de^3({\bf r}),
\ee
which is consistent with (12) and (23) for static force or potential produced by a relativistic particle.

The relativistic static equation (25) leads to a Coulomb-like potential energy $g_{\ell}L_0({\bf r})=- {g^2_\ell(E_{\nu}/m_{\nu})}/{4\pi r}$.  Thus, the attractive Lee-Yang force $F^{LY}$ between a 
non-relativistic electron $(i.e., E_e/m_e \approx 1)$ and a relativistic anti-e-neutrino with opposite leptonic charges and with large energy $(E_{\nu}/m_{\nu}) >>1$ is given by
\be
  F^{LY }=- g_\ell \frac{d L_0}{dr}=\frac{-g^2_\ell(E_{\nu}/m_{\nu})}{4\pi r^2}.
\ee
We shall use this result to discuss the anti-e-neutrino dark matter.

\section{Cosmic Lee-Yang force, relativistic gravitational force and flat rotation curves}

Let us now demonstrate that the cosmic Lee-Yang force $F^{LY}$ in (26) due to leptonic charges can result in a flat rotation curve.  Suppose a spiral galaxy has visible mass $M$. Because the galaxy has approximately equal numbers of protons and electrons and the mass of the galaxy is principally due to its protons\footnote{Let us ignore the neutrons in the estimation because 75\% of cosmic mass is hydrogen.}, the galaxy contains approximately $M/m_p$ electrons where $m_p$ is the mass of a proton. Hence, its leptonic charge is given by $g_\ell M/m_p$. Because the model under discussion assumes the dominance of neutron decays in the beginning of the universe, the number of anti-e-neutrinos should be roughly the same as the number of electrons or protons in the galaxy. Thus, the anti-e-neutrinos in a spiral galaxy contribute leptonic charges of $-g_\ell M/m_p$.

Based on these considerations and the results (22) and (26), we discuss the role played by relativistic anti-e-neutrinos (i.e., $E_\nu >> m_\nu < 1.1 \ eV$ \ \cite{5}) in a simplified model of the observed flat rotation curves of spiral galaxies.  For simplicity, suppose $M$ is the mass of a galaxy within a sphere of radius r (assuming spherical symmetry) and a suitably located star (with mass $m$) from the center of the galaxy with a distance $r$.  For the inverse-square force, we have the combined Lee-Yang and gravitational forces,
\be
F_{com}= F_{YM} + F^{LY}
\ee
$$
F_{YM}=-\frac{GmME_{\nu}}{r^2 m_p}(1+\b^2), \ \ \ \ \    F^{LY} =-\frac{g_{\ell}^2 m M E_\nu}{4\pi r^2 m^2_p m_{\nu}},
$$
where we have used (26) and the first equation in (22) for concreteness, $M/m_p$ is the approximate number of anti-e-neutrinos (or electrons or protons) in the galaxy, and $m/m_p$, where $m_p=0.94 GeV$, is approximately the number of electrons in the star with mass $m$.  Equation (27) involves two unknown quantities, $g_{\ell}$ and $E_\nu$.  Since there is no data of energy spectrum for cosmic anti-e-neutrino, let us assume the energy $E_\nu$ to be of the order of magnitude of MeV in the expanding universe to estimate  leptonic charges.\footnote{Since both electrons and anti-e-neutrinos in the observable universe are assumed to be the product of neutron decays in this model of dark matter, let us assume that their average kinetic energies are of the order of magnitude, say, $~ MeV$.  The purpose is to get a very rough estimate of the leptonic charge $g_{\ell}$.  Different values of $E_{\nu}$ will lead to different values of $g_{\ell}$, as shown in Tables 1 and 2.}

In the conventional interpretation of the flat galactic rotation curve,\cite{13}  the mass of the dark matter in a galaxy is taken to be $M_D \approx N_D M$ where $N_D$ is estimated to be on the order of 5 to 60\cite{14,15}.  Since this dark matter is assumed to interact only gravitationally, the gravitational force between the galactic dark matter with mass $M_D$ (with $N_D=5$ for concreteness) and a suitably located object with mass $m$ is 
\be
F^{grav}= -\frac{G m M_D}{ r^2},  \ \ \ \ \ \ \ \ \   M_D \approx 5M.
\ee
Thus, in order to produce the observed flat rotation curve, the combine forces in (27) must be equal to the corresponding effective gravitational force (28), 
$$
 \frac{GmME_{\nu}}{r^2 m_p}(1+\b^2)+ \left(\frac{g^2_\ell m M}{m^2_p} \right)\left(\frac{E_\nu}{m_\nu}\right) \frac{1}{4\pi  r^2} \approx \frac{G m ( 5M)}{ r^2}. 
 $$
 \be
 (1+\b^2) \frac{E_\nu}{m_p} + \frac{g_{\ell}^2 E_\nu}{4\pi G m^2_p m_\nu} \approx 5, \ \ \ \ \  M_D \approx 5M.
 \ee
 
 So far, we do not have data for the cosmic neutrino energy $E_\nu$ in (29).  Since it was created in the neutron decay in the beginning of the universe, let us assume a value,  $E_\nu/m_\nu\approx 10^6$, to estimate roughly the leptonic charge $g_{\ell}$.   Based on (29) with $c=\hbar=1$, $m_p= 0.94GeV$, $G=6.7\times 10^{-39} GeV^{-2}$, $E_\nu/m_\nu \approx 10^6$, and $\b \approx 1$ for the relativistic velocity of $\ov{\nu}_e$, we estimate that 
\be
  g_{\ell} \approx  7 \times 10^{-21}, \ \ \ \ \ for \ \    M_D\approx  5M.
\ee
$$
g_{\ell} \approx  2 \times 10^{-21},  \ \ \ \ \ for \ \    M_D\approx  60M.
$$

The value for the leptonic charge $g_{\ell}$ in (30) is extremely small.  The ratio of the leptonic  and the electromagnetic coupling strengths is roughly
\be
{(g^2_\ell/4 \pi)}/{(e^2/4\pi)}\approx 5 \times 10^{-41}, \ \ \ \ \   e^2/4\pi = 7.3 \times10^{-3}
\ee
 The ratio of ${(g^2_\ell/4 \pi)}$ and $(G m^2_p)$ is roughly
  \be
  {(g^2_\ell/4 \pi)}/{(G m^2_p)}\approx 6 \times 10^{-6},  \ \ \ \ \ \     G m^2_p \approx  6\times 10^{-39}.
   \ee
  
  In the estimations of the leptonic charge $g_{\ell}$ based on (29), the contribution of the gravitational force of neutrinos is small for $E_\nu  < 10^9 m_\nu$.  Only for the relativistic $\ov{\nu}_e$ with energy $E_\nu > 10^9 m_\nu $, the contribution of the gravitational force should also be taken into account.   In Table 1 with $M_D=5M$, when energy $E_\nu > 10^{10} m_\nu $, the gravitational force in the first term of (29) dominates so that there is no positive solution for the square $g_{\ell}^2$ of the leptonic charge.  In the following discussion in section 8, precision Cavendish experiments could measure $g^2_{\ell}/(4\pi G m^2_p)$ up to roughly $10^{-5}$.
  
   In conventional models of dark matter, the flat rotation curves is related to the property that  the dark matter has only gravitational force and is distributed with a density roughly proportional to $1/r^2$.  In the HEN model, the enhanced gravity of anti-e-neutrinos can contribute to the flat rotation curves if $E_\nu > 10^9 m_\nu$.  When $E_\nu$ is smaller than, say, $10 m_\nu$, the gravitational force of $\ov{\nu}_e$ is negligible, as one can see from (29).  In this case, the flat rotation curve is due to leptonic Lee-Yang force of anti-e-neutrinos, whose density distribution is roughly $1/r^2$ with a suitable coupling constant $g_{\ell}$ because the Lee-Yang force is proportional to $ r^{-2}$.
 
\bigskip
 
\bigskip

 \begin{center}
{\bf Table 1.  Estimations of $g^2_{\ell}/(4\pi G m^2_p)$ with (18) vs. energy $E_{\nu}$}
\end{center}
$$ eq. \ (29):  \ \ \ \ (1+\b^2)(E_\nu/m_p) + [g_{\ell}^2(E_\nu/m_\nu)/(4\pi G m^2_p)] \approx 5 \ \ \ or \ \ 60 $$
  ------------------------------------------------------------------------------------------------
$$ 
E_\nu: \ \ \ \ \ \ \ \  10^3m_\nu \ \ \ \ \ \ \ \   10^6m_\nu \ \ \ \ \ \ \ \  10^9m_\nu \ \ \ \ \ \ \ \ 10^{10}m_\nu \ \ \ \ \ \ \ \ 10^{11}m_\nu 
$$
  ------------------------------------------------------------------------------------------------
$$ 
\frac{g^2_{\ell}}{4\pi G m^2_p}:  \ \ \ \ \      5\times 10^{-3} \ \ \ \     5\times 10^{-6} \ \ \ \ \   3\times 10^{-9} \ \ \ \ \ \ \  no\ solu. \ \ \ \   no\ solu. 
$$
$$ 
({M_D}=5 M) \ \ \   
 $$
   ------------------------------------------------------------------------------------------------
$$ 
\frac{g^2_{\ell}}{4\pi G m^2_p}:  \ \ \ \ \        6\times 10^{-2}  \ \ \ \ \ \     6\times 10^{-5} \ \ \ \ \   5.8\times 10^{-9}\  \ \ \ \ \ \ \ \  4\times 10^{-9}  \ \ \ \ \ \ \   no\ solu. 
$$
$$ 
({M_D}=60 M) \ \ \   
$$
 {\bf ********************************************************}
 
   \bigskip
   
   \bigskip
   
 \begin{center}
{\bf Table 2.   Estimations of $g^2_{\ell}/(4\pi G m^2_p)$ with (19)\footnote{It corresponds the second equation in (23).} vs.  energy $E_{\nu}$}
$$  (1+\b^2)(E_\nu/m_p)^2 + [g_{\ell}^2/(4\pi G m^2_p)](E_{\nu}/ m_\nu) \approx 5 \ \ \ or \ \ 60 $$
\end{center}
  ------------------------------------------------------------------------------------------------
$$ 
E_\nu: \ \ \ \ \ \ \ \ \ \ \ \ \ 10^2m_\nu \ \ \ \ \ \ \ \ \ \ \ \ \ \ \  10^4m_\nu \ \ \ \ \ \ \ \ \ \ \ \ \ \  10^5m_\nu \ \ \ \ \ \ \ \ \ \ \ \ \ \ \ 10^{6}m_\nu 
$$
  ------------------------------------------------------------------------------------------------
$$ 
\frac{g^2_{\ell}}{4\pi G m^2_p}:  \ \ \ \ \ \ \ \     5\times 10^{-2} \ \ \ \ \ \ \ \ \ \  4.8\times 10^{-4} \ \ \ \ \ \ \ \ \ \ \ \  no\ solu. \ \ \ \ \ \ \ \ \ \  no\ solu. 
$$
$$ 
({M_D}=5 M) \ \ \   
 $$
   ------------------------------------------------------------------------------------------------
$$ 
\frac{g^2_{\ell}}{4\pi G m^2_p}:  \ \ \ \ \ \ \ \ \        60\times 10^{-2}  \ \ \ \ \ \ \ \ \ \ \ \ \ \    60\times 10^{-4} \ \ \ \ \ \ \ \ \ \ \ \ \ \  4\times 10^{-4}\    \ \ \ \ \ \ \ \ \ \ \ \ \ \ \   no\ solu. 
$$
$$ 
({M_D}=60 M) \ \ \   
$$
 {\bf ********************************************************}

 \section{Alpha magnetic spectrometer experiment and anti-e-neutrino dark matter}
 
  Result (32) reveals that the new leptonic Lee-Yang force between two electrons  is much weaker  than the gravitational force and, hence, it cannot be detected in high energy laboratories.  Only when one considers a physical system involving extremely large number of leptonic charges can one detect the effect of the Lee-Yang force.  Thus, besides observations of rotation curves, which involve a galaxy's worth of leptons, it is in general very difficult to test the present model of anti-e-neutrino dark model by other independent experiments. 
  
Fortunately,  anti-e-neutrinos $\ov{\nu}_e$ also interact via the usual weak interactions\footnote{This interaction is effectively characterized by the Fermi constant\cite{2} $G_F=1.17\times 10^{-5} GeV^{-2}$, which is much larger than Newton's constant $G=6.7\times 10^{-39} GeV^{-2}$.} in unified electroweak theory.  This weak interaction force has a much stronger coupling strength\cite{2,3} ($\approx  G_F m_p^2=10^{-5}$) than  that of the leptonic Lee-Yang force or the gravitational force.  Therefore, it is possible for  the dark matter, i.e., the ubiquitous
 anti-e-neutrino to interact with cosmic-ray protons to produce positrons through the electroweak process,
 \be
 \ov{\nu}_e + p^+ \to e^+ + n.
 \ee 
 Normally, without the ubiquitous anti-e-neutrinos, one would not expect positrons to be created by cosmic rays in any significant quantity.  How then, can one look for interactions between anti-e-neutrinos ($\ov{\nu}_e$) and protons as in (33)?
 
 It turns out that the Alpha Magnetic Spectrometer on the International Space Station has detected an intriguing excess of cosmic-ray positrons over what would normally expected\cite{16}. 
   Thus, this model in which the effects of an apparent ``dark matter" are instead the result of leptonic Lee-Yang forces generated by the interactions of anti-e-neutrinos and electrons can be tested by the Alpha Magnetic Spectrometer (AMS) experiment.

Let us briefly discuss the weak process (33) based on dimensional considerations and order-of-magnitude estimations. The total cross section of the process (33), which summed over all final states, is denoted by $\s(\ov{\nu}_e p^+)$.  The amplitude of the weak process (33) should be proportion to the Fermi constant  $G_F=1.17\times 10^{-5} GeV^{-2}$. Its total cross section $\s(\ov{\nu}_e p^+)$ must have the form\cite{2} 
\be
\s(\ov{\nu}_e p^+) = G^2_F f(s, m_p), \ \ \ \      G_F \approx 10^{-5} GeV^{-2},
\ee
where $s$ is the square of the center-of-mass energy.  At high energies $s >> m^2_p$, the model of anti-e-neutrino dark matter predicts
\be
\s(\ov{\nu}_e p^+) \approx G^2_F s.
\ee
We may compare the total cross section (35) with neutrino-nucleon scattering in high energy laboratories\cite{2},
\be
\s_{exp} \approx 10^{-38}\left( \frac{E_{\nu}}{m_p}\right) cm^2,
\ee
where the nucleon is at rest.   

If the prediction (35) of the HEN model is tested and confirmed by the AMS experiment, it would support the interpretation of the dark matter as high energy anti-e-neutrinos and the postulate of the universal gauge symmetry.

\section{Precision Cavendish experiments to test the leptonic Lee-Yang force} 
Another possible test of the cosmic Lee-Yang force involves the force between two macroscopic bodies in a precision Cavendish experiments.\cite{17}  Consider the Lee-Yang force $F^{LY}$ between a pendulum ball with mass $m_{p}$ and a source ball with mass $m_s$  in a modern precision Cavendish experiment.  For simplicity, we ignore the gravitational force due to the electrons and consider only the leptonic Lee-Yang force between the electrons of the two balls, which may be made of different materials. Suppose the pendulum ball with mass $m_{p}$ is composed of an element with atomic number $Z_p$ and atomic mass $A_p$, and the source ball with mass $m_s$ is composed of an element with atomic number $Z_s$ and atomic mass $A_s$. 

The total number of the electrons in the source mass $m_s$ is the number of atoms in it, $m_s/(A_s a_u)$, times its atomic number $Z_s$.  Thus, the total number of leptonic charges carried by electrons in the source mass is  $g_{\ell s}= g_{\ell}m_s Z_s/(A_s a_u)$, where $a_u$ is the atomic mass unit.  Similarly, the total leptonic charges in a pendulum ball with mass $m_{p}$ is  $g_{\ell p}=g_{\ell}m_{p} Z_p/(A_p a_u)$.  We note that the  observed  force $F_{ob}$ between the source mass $m_s$ and the pendulum ball with mass $m_p$ actually includes both the repulsive Lee-Yang force $F^{LY}= {g_{\ell p}g_{\ell s}}/{4\pi r^2}$ and the true attractive gravitational force $F_g$:
\be
 F_{ob}=  -\frac{G_{ob} m_p m_s}{r^2}=F_g + F^{LY},   
\ee
$$
 F_{g}=-\frac{G_g m_p m_s}{r^2}, \ \ \  F^{LY}=  \left(\frac{ m_p Z_p}{A_p } \right)\left(\frac{ m_s Z_s}{A_s }\right)  \frac{g_\ell^2}{4\pi  r^2 a^2_u}, 
$$
$$
a_u=0.93149 \ GeV = 1.66054\times 10^{-27} \ kg, \ \ \ \ \ \     c=\hbar=1.
$$
where the observed value $G_{ob}$ involves both contributions from the leptonic charges and from  the true gravitational constant $G_g (\ne G_{ob})$.  Based on experimental data, we know that the magnitude of the lepton force is small than that of the gravitational force because the first 3 numbers of the observed Newtonian constant (i.e., 6.67U) are the same in all reliable experiments.   
  
 As a concrete example, the precision `observation 1' gives the Newtonian constant $G_{ob1}=6.67559U$,\cite{18,17} where $U\equiv 10^{-11} m^3 kg^{-1} s^{-2}=1.0052\times 10^{-39} GeV^{-2}$.  It is based on the Cavendish type apparatus 
  with copper (Cu) pendulum ball of mass $m_{p1}$ and copper source mass $m_{s1}$.  We have 
 \be
 (g_{\ell p}g_{\ell s})_{Cu}=  \left(\frac{g^2_\ell m_{p1} m_{s1}}{ a^2_u} \right)\left(\frac{Z_{p1} Z_{s1}}{A_{p1} A_{s1}}\right)_{Cu},
\ee 
$$
\left(\frac{Z_{p1} Z_{s1}}{A_{p1} A_{s1}}\right)_{Cu}=\left(\frac{29}{63.546}\right)^2= 0.20824.
$$

For `observation 2', we have $G_{ob2}=6.67418U,$\cite{19} the pendulum is quartz $(Si O_2)$ with mass $m_{p2}$ and the source ball is made of iron with mass $m_{s2}$.   Now, we have 
 \be
 (g_{\ell p})_{_{Si O_2}}(g_{\ell s})_{_{Fe}}=  \left(\frac{g^2_\ell m_{p2} m_{s2}}{ a^2_u} \right)\left(\frac{Z_{p2}}{A_{p2}}\right)_{SiO_2}\left(\frac{Z_{s2}}{ A_{s2}}\right)_{Fe},
\ee 
$$
\left(\frac{Z_{p2}}{A_{p2}}\right)_{Si O_2}\left(\frac{Z_{s2}}{ A_{s2}}\right)_{Fe}
=\left(\frac{(14+16)}{(28+32)}\right)\left(\frac{26}{55.845}\right)= 0.23279 .
$$

Let us consider the difference of the following two quantities,
$$
\frac{F_{ob1}}{(m_{p1} m_{s1})} - \frac{F_{ob2}}{(m_{p2} m_{s2})}=
\frac{-G_{ob1}m_{p1} m_{s1}}{r^2(m_{p1} m_{s1})} + \frac{G_{ob2}m_{p1} m_{s2}}{r^2(m_{p2} m_{s2})}.
$$
\be
=\frac{g^2_\ell }{4\pi r^2 a^2_u}\left[ \left(\frac{Z_{p1} Z_{s1}}{A_{p1} A_{s1}}\right)_{Cu}  \right.
- \left. \left(\frac{Z_{p2}}{A_{p2}}\right)_{Si O_2}\left(\frac{Z_{s2}}{ A_{s2}}\right)_{Fe} \right],     
\ee
in which the true gravitational forces in $F_{ob1}/(m_{p1} m_{s1} )$ (i.e., $G_{g}/r^2$)  and in $F_{ob2}/(m_{p1} m_{s2} )$ ( i.e., $G_g/r^2$)  cancel.

The strength of the leptonic charge $g^2_{\ell}/4\pi$ can be estimated from the observed difference of forces in (40).  We obtain
\be
\frac{g^2_\ell }{4\pi}= - a^2_u \left[G_{ob1}-G_{ob2}\right] Q \approx 5 \times 10^{-43},
\ee
$$
Q\equiv \left[ \left(\frac{Z_{p1} Z_{s1}}{A_{p1} A_{s1}}\right)_{Cu} -  \left(\frac{Z_{p2}}{A_{p2}}\right)_{SiO_2}\left(\frac{Z_{s2}}{ A_{s2}}\right)_{Fe}\right]^{-1}.   
$$
where we have used (38) and (39) and the experimental values of $G_{ob1}=6.67559U$ and $G_{ob2}=6.67418U$.\cite{18,19}
\newpage
  \bigskip
 \begin{center}
{\bf Table 3.   Estimations of $g^2_{\ell}/(4\pi )$ with equation (41).}
\end{center}
$$  -------------------------------------$$
$$  G_{ob1}: \ \ \ \ \ \ \ \ \ \  6.67559U \ \ \ \ \ \ \ \ \ \ \ \ \ \ \  6.67559U \ \ \ \ \ \ \ \ \ \ \ \ \ \ \ \  6.67559U \ \ \ \ \ \ \ \ \ $$
$$  --------------------------------------$$
$$ 
G_{ob2}: \ \ \ \ \ \ \ \ \ \   6.67418U \ \ \ \ \ \ \ \ \ \ \ \ \ \   6.67349U \ \ \ \ \ \ \ \ \ \ \ \ \ \ \ \ \    6.67191U  \ \ \ \ \ \ \
$$
$$  --------------------------------------$$
$$
 \ \ \ \ \ \ \ Q :  \ \ \ \ \ \ \ \ \ \ \    Q_1= -0.02455 \ \ \ \ \ \ \ \ \ \ \ \ \  Q_1= -0.02455 \ \ \ \ \ \ \ \ \ \ \ \ \ \ \  Q_2= +0.0339 \ \ \ \ \ \ \ \ \ \ \
$$
$$ \ \ \ \ \ Q_1 \equiv 1/[ \left({Z_{p1} Z_{s1}}/{A_{p1} A_{s1}}\right)_{Cu} -  \left({Z_{p2}}/{A_{p2}}\right)_{SiO_2}\left({Z_{s2}}/{ A_{s2}}\right)_{Fe}].    \ \ \ \ \ \ \ \ \ \ \ \ \ \ \ \ \ \       $$
$$ \ \ \ \ \ \ \  Q_2 \equiv 1/[ \left({Z_{p1} Z_{s1}}/{A_{p1} A_{s1}}\right)_{Cu} -  \left({Z_{p2}}/{A_{p2}}\right)_{W}\left({Z_{s2}}/{ A_{s2}}\right)_{Rb}].   \   W=tungsten        $$
$$  ----------------------------------------$$
$$
g^2_{\ell}/(4\pi):  \ \ \ \ \ \ \ \ \ \ \ \ \ \      5\times10^{-41} \ \ \ \ \ \ \ \ \ \ \ \ \ \ \ \ \   7.4\times 10^{-41} \ \ \ \ \ \ \ \ \ \ \ \ \ \ \ \ \ \ \   no\ \  positive \ \ solu.  \ \ \ \ \ \  $$
$$  ----------------------------------------$$

\section{Discussions}

The AMS experiment also collects data of antimatter.\footnote{The AMS experiment in the International Space Station is a big collaboration involving 500 scientists from 16 countries. }  In this connection, we note that particle-cosmology with CPT invariance\cite{2} suggests a big picture of the universe with matter half-universe and antimatter half-universe.  CPT invariance implies a maximum symmetry between particle and antiparticle, e.g., lifetime and mass equalities between particles and antiparticles\cite{2}. According to the Big Jets model,\cite{7} one may picture two jets as two gigantic fireballs expanding and moving away in opposite directions in the beginning of the universe. Each jet has extremely large number of quarks and antiquarks.  After complicated processes of  collision, confinement, strong decay, annihilation, etc., if one fireball happened to be dominated by particles, then the other fireball must be dominated by corresponding antiparticles.  The evolution in each fireball after the neutron (or anti-neutron) decay is similar to that in a hot big bang.   Thus, the Big Jets model with CPT invariance\cite{7} predicts that the antimatter half-universe has electron-neutrinos `anti-dark-matter' and  should be far away from our matter half-universe.  Hence, we cannot detect antimatter in our matter half-universe.  Such a separation of matter from antimatter in two separate half-universes appears to be consistent with the result (i.e., the absence of antimatter) of AMS experiment\cite{16}. 

\subsection{Features of HEN model and related phenomena}

This high energy neutrino (HEN) model demonstrates that anti-e-neutrinos are a viable dark matter candidate and deserve further investigation.  This model represents an alternative approach to understanding dark matter phenomena that does not depend on the existence of as-yet undiscovered particles and other speculations. The features of the HEN model are as follows:

 (i) It provides an understanding of the flat galactic rotation curves without drastically increasing the total observed mass of the universe by a factor of 5 to 60, in contrast to all other interpretations of dark matter. 
 
 (ii) It is based on the well-established principle of gauge symmetry for quantum Yang-Mills gravity.  (In particular, at the super-macroscopic limit, Yang-Mills gravity and the cosmological principle lead to a Hamilton-Jacobi type equation for the motion of galaxies.  Its solutions are consistent with Hubble's Law and cosmic redshifts.)
 
  (iii) It is based on particle-cosmology in a flat space-time with CPT invariance, implying the conservation of leptonic and baryonic charge, as well as an electrically neutral universe. 
    
  (iv) The HEN model assumes that high energy anti-e-neutrinos were produced through a dominant neutron decay process, $n\to p^+ + e^- + \ov{\nu}_e$, within the first 1000 seconds after the creation of quarks and antiquarks, at the beginning of the universe.  This assumption is consistent with experiments that there are almost equal number of protons and electrons in the observable cosmos.
  
 (v) The model predicts that the interaction between these ubiquitous anti-e-neutrinos and cosmic-ray protons leads to an excess of cosmic-ray positrons, which have been detected by the Alpha Magnetic Spectrometer experiment.  An excess of positrons in cosmic rays was also found by the HEAT balloon experiment by Barwick et al and PAMELA.\cite{16} 
 
Investigations of a double galaxy cluster (bullet cluster) give us a clear property that dark matter (DM) can only have extremely small force, similar to gravity, with itself or with baryons.  We know that the galaxies in this cluster are mainly  grouped to two distinct sub-clusters.  There is hot gas which emits x-rays and is concentrated between the sub-clusters. This phenomenon suggests that the two clusters of galaxies have collided.  Only the two clouds of hot gas that previously accompanied  have collided and, hence, remained closer to the center of the double cluster.  It was found that most matter in the double galaxy cluster is not associated with the hot gas, but like the galaxies form two sub clusters that have passed through each other without appreciate interaction.\cite{20}  This appears to be consistent with the HEN model of dark matter with a suitable energy spectrum and density.   Because thus far, there is no data regarding the energy spectrum of cosmic anti-e-neutrinos, we can only roughly estimate the forces and coupling constants involved in the HEN model.   
 
There are other phenomena such as the structure hierarchy of the universe, CMB, etc.   What we know about dark matter (DM) and the disk of galaxies is that dark matter particles must not have significant interaction with radiation because DM has not lost kinetic energy sufficiently to relax into the disk of galaxies, in contrast to baryon matter.\cite{20}  The HEN model appears to be consistent with this property because the coupling constants associated with neutrinos are extremely weak, as shown in (32) and Tables 1 and 2.  These phenomena need to be further studied within the framework of particle-cosmology with Yang-Mills gravity in flat space-time.  However, journal articles have length limitations and such phenomena are beyond the scope of the present paper. 

One may wonder about inflation (or the flatness problem) in the present particle-cosmology.
It is interesting to note that the particle-cosmology with Yang-Mills gravity in  flat space-time provides the simplest solution to the flatness problem. That is, we are in a universe with flat space-time, in which $\rho$ is always precisely equal to $\rho_{crit}.$  A conventional solution of the flatness problem is given by inflationary theories.\cite{20} 

Moreover, the framework of particle-cosmology with general Yang-Mills symmetries provides a new perspective of dark energy.  The general $U_1$ gauge symmetry for baryon charges can lead to a linear repulsive force with an extremely small coupling constant. As the universe evolved, there came a time when the average distance between galaxies exceeded a critical value $R_c$, leading to a situation in which the linear repulsive baryonic force is larger than the gravitational attractive force between two galaxies.  This could cause the observed late-time accelerated cosmic expansion. However, we refer to references for it is beyond the scope of this paper.\cite{20,21}

Generally speaking, it is too early to reach any definite conclusion from previous consideration.  For example, it may be possible for the dominant neutron decays to occur at  a different time.  However, if there is a beginning of the universe with the creation of the strongly interacting quarks and anti-quarks, then there must be extremely complicated interactions, including decays after the formation of hadron states.  The stable and non-strong interacting particles such as electrons and neutrinos cannot come from from the direct production processes with the same number as protons and neutrons.  It seems likely that stable particles will arise from decay processes before nucleosynthesis.  We hope that in the future there will be more data to test the HEN model.
 
 \subsection{Mathematical aspects of Yang-Mills gravity and Einstein gravity}

 Foundational physics is defined as the invariance of action under local gauge transformations.  Different choices of gauge groups lead to different field theories.  If one chooses internal gauge group  $SU_3$, one has quantum chromodynamics.  If one chooses external space-time translation group, one has quantum Yang-Mills gravity\cite{4};  and if one chooses general coordinate transformation group, one has classical Einstein gravity.  It turns out that these two different  formulations of gravity are based on formally the same space-time transformations:
\be
x^\mu \to x'^\mu= x^\mu + \Ld^\mu(x), \ \ \  x\equiv x^\ld,
\ee 
where $\Ld^\mu(x)$ is an arbitrary infinitesimal vector function, which can be identified with the vector field in the Lie derivative $\mathcal{L}_\Ld  $ of arbitrary tensors. 
Let us briefly consider two different viewpoints of the space-time transformations (42):

(A) {\em Yang-Mills gravity}. \ \ \ One interprets $\Ld^\mu(x)$ in (42) as local translations in flat space-time.  The Yang-Mills gauge covariant derivative $\De_\mu$ involves the representations of the generators $p_\mu = i\p_\mu$ of  the space-time translation ($T_4$) group, 
\be
\D_\mu \psi= \p_\mu \psi + ig\phi_\mu^\nu(x) p_\nu \psi = [\p_\mu  -g\phi_\mu^\nu(x) \p_\nu]\psi, \ \ \ \ (c=\hbar=1),
\ee
which dictates the universal coupling between gravitational symmetric tensor field $\phi^{\mu\nu}=\phi^{\nu\mu}$ and fermion fields $\psi$ of, say, leptons or quarks.  At the same time, (42) can also be considered as the coordinates transformations in inertial frame with Minkowsky metric tensor $\e_{\mu\nu}=(1,-1,-1,-1)$ (or in non-inertial frames with Poincar\'e metric tensor $P_{\mu\nu}$, which reduces to the Minkowsky metric tensor in the limit of zero acceleration\cite{4}).  In Yang-Mills gravity, the space-time coordinates $x^\mu$ have the usual operational meanings, as required by quantum field theories with Poincar\'e invariance\cite{4}.

(B) {\em Einstein gravity}.  \ \ \ \ One interprets (42) as general coordinate transformations, which include  all one-to-one and twice-differentiable transformations of the coordinates.  This interpretation follows from the principle of general coordinate invariance, i.e., the laws of physics should be invariant under the general coordinate transformations (42).  This principle clearly implies that space-time coordinates can be given  arbitrary labels and, hence, do not have the usual operational meaning for space and time coordinates.\cite{22}  One has Riemannian covariant derivative in curved space-times, e.g.,
\be
D_\mu V^\nu(x)=\p_\mu V^\nu(x) +\G_{\mu\ld}^\nu(x) V^\ld(x),
\ee
where $\G^\nu_{\mu\ld}(x)$ is torsionless Levi-Civita connection.  The vector field $V^\nu(x)$ in the expression (44) can be generalized to arbitrary tensor fields.  The coordinate $x^\mu$ in (44) is local in Riemannian manifold and has no meaning in itself\cite{23}.  

  It is interesting to note that both covariant derivatives (43) and (44) can be considered as two special cases of covariant derivatives in the theory of Lie derivative in coordinate expressions\cite{24}.  The $T_4$ covariant derivative (43) is new in the sense that it is related to the `new interpretation' of the transformations (42) as the flat space-time translations $T_4$ group, while the usual covariant derivative in the theory of Lie derivative involves any symmetric (or torsion-free) covariant derivative.  The expression (44) is a special case that involves the Levi-Civita connection\cite{24}.
  
 We note that the $T_4$ space-time translational gauge transformations of arbitrary tensors with infinitesimal vector gauge function $\Ld^\mu(x)$ in Yang-Mills gravity is exactly the same as the Lie derivatives  $\mathcal{L}_\Ld  $ of arbitrary tensors in the coordinate expressions in flat space-time\cite{4}.  The arbitrary gauge function $\Ld^\mu(x)$ in (42) can be identified with the vector function in the Lie derivatives  $\mathcal{L}_\Ld  $.  Furthermore, the $T_4$ gauge invariance of an action in Yang-Mills gravity is the same as the vanishing of the Lie derivative of the action.  H. Cartan's formula\cite{23}\footnote{H. Cartan is the son of E. Cartan.} facilitates the calculation of the change of the volume $Wd^4x$ and the invariance of the action (1) in Yang-Mills gravity under the $T_4$ gauge transformations\cite{4}.   In this sense, the theory of Lie derivatives (in coordinate expressions) is the mathematical basis of Yang-Mills gravity.\footnote{Historically, Pauli appears to have been the first to discuss a new variation $\d^* a^\mu =a'^\mu (x)-a^\mu(x)$ for all tensors in his book on relativity (p.66) published in 1921.  \'Slebodzi\'ski introduced in 1931  a new differential operator for all tensors in his discussion of Hamilton's equations.  Later, it was named the Lie derivative.  Cf.  ref. 4, pp. 108-111.}
 
 Apart from conceptual problems for particle-cosmology when it is based on general relativity, as discussed in sec. 2, there are also basic problems related to experiments and observations.  Wigner wrote: `The basic premise of this theory [the general theory of relativity] is that coordinates are only auxiliary quantities which can be given arbitrary values for every event.  Hence, the measurement of position, that is, of the space coordinates, is certainly not a significant measurement if the postulates of the general theory are adopted: the coordinates can be given any value one wants.  The same holds for momenta.  Most of us have struggled with the problem of how, under these premises, the general theory of relativity can make meaningful statements and predictions at all'\cite{22}.

Finally, based on particle-cosmology with fundamental CPT invariance\cite{2} and universal gauge symmetry,\cite{4} we suggest that the universe began with two Big Jets\cite{6} with quarks and antiquarks and evolved to two half-universes, i.e., 3K matter and antimatter blackbodies.  The matter half-universe consists of four matter particles, neutrons, protons, electrons and anti-electron-neutrinos (i.e., dark matter), together with gauge bosons.  The antimatter half-universe consists of the corresponding antiparticles, according to CPT invariance.  We demonstrate that the  ubiquitous and unobservable anti-e-neutrinos plays a role in effects attributed to dark matter because the additional leptonic Lee-Yang interaction in which anti-e-neutrinos  participate can also contribute to observed flat galactic rotation curves.  The interaction of ubiquitous anti-e-neutrinos and cosmic-ray protons leads to an excess of cosmic-ray positrons, which have been detected by the Alpha Magnetic Spectrometer experiment and other experiments\cite{25,26}.  If the cosmic-ray process (33) with the total cross section (35) involving  anti-e-neutrinos in the present model is confirmed by the AMS experiment, it would be evidence supporting this model of anti-e-neutrino dark matter.  Such AMS result could also support particle-cosmology with Yang-Mills gravity and the universal principle of gauge symmetry.  

The work was supported in part by Jing Shin Research fund and Prof. Leung Memorial Fund of the UMassD Foundation.  The author would like to thank Y. Zhu for useful discussions.  He also thanks the Reviewer1 for useful comments.

\bibliographystyle{unsrt}

\end{document}